\documentstyle[11pt]{article}

\begin{document}

\thispagestyle{empty}

\mbox{}

{\raggedleft

WU-AP/50/95 \\
SUSSEX-AST-95/9-1 \\
hep-th/9509074 \\}

\vspace{7mm}

\begin{center}  {\Large\bf\expandafter{
           Tree-level String Cosmology }} \end{center}

\vspace{5mm}

\begin{center}
{\large Richard Easther} \footnote{ easther@cfi.waseda.ac.jp}\\
{\large Kei-ichi Maeda} \footnote{ maeda@cfi.waseda.ac.jp} \\
Department of Physics, School of Science and Technology, Waseda
University, \\ 3-4-1 Okubo, Shinjuku-ku, Tokyo, Japan.

\bigskip
{\large David Wands}\footnote{  d.wands@sussex.ac.uk} \\Astronomy
Centre,  School of Mathematical \& Physical Sciences, \\
University of Sussex, Brighton BN1 9QH., U.~K.  \end{center}

\vspace{1cm}

\setcounter{footnote}{0}
\setcounter{page}{0}
\section*{Abstract}

In this paper we examine the classical evolution of a cosmological model
derived from the low-energy tree-level limit of a generic string theory.
The action contains the metric, dilaton, central charge and an
antisymmetric tensor field. We show that with a homogeneous and
isotropic metric, allowing spatial curvature, there is a formal
equivalence between this system and a scalar field minimally coupled to
Einstein gravity in a spatially flat metric. We refer to this system as
the shifted frame and using it we describe the full range of
cosmological evolution that this model can exhibit.  We show that
generic solutions begin (or end) with a singularity.  As the system
approaches a singularity the dilaton becomes becomes large and loop
corrections will become important.

\medskip
\vfill

PACS: 98.80.Cq, 11.25.-w, 04.50.+h
\vfill
\mbox{}

\newpage
\section{Introduction}

String theory is the most promising candidate for the unification of
gravity with the other fundamental forces of nature. However, string
theory is most likely to cause significant modifications to classical
general relativity near the Planck scale, which is far beyond the
range of direct terrestrial experimentation. Since these energy scales
are typically associated with the Big Bang it is natural to view
cosmology as a laboratory for testing string theoretic modifications
to gravity.  Conversely, cosmologists can hope that if string theory
does provide a deeper understanding of gravitational physics than
general relativity then some of the ``standard'' problems of
conventional cosmology will be resolved by string theory.
Consequently, the cosmological dynamics of superstring theories are
the subject of intense scrutiny. Typically, one proceeds by perturbatively
expanding the full superstring theory and extracting a ``low energy''
Lagrangian that contains Einstein gravity and the lowest order
corrections from string theory. For this approach to be valid we
need to  restrict our attention to sub-Planckian scales where quantum
effects can be ignored and the higher order terms in the perturbative
expansion do not contribute significantly.

We consider the tree-level action
\cite{FradkinET1985a,CallanET1985a,Lovelace1986a} which contains
contributions from the metric, the dilaton, the central charge and an
antisymmetric tensor field. We assume that the background spacetime is
four dimensional and that it, and the fields defined upon it, are
homogeneous and isotropic. Further, we assume that degrees of freedom
associated with any compactified metric can be ignored.  Such
cosmologies have received considerable study,
\cite{%
BailinET1985b,Maeda1986Pa,Maeda1987a,%
AntoniadisET1988a,AntoniadisET1989a,LiddleET1989a,%
Mueller1990a,AntoniadisET1991b,Veneziano1991a,Tseytlin1991a,%
CasasET1991b,GarciaBellidoET1992b,%
Tseytlin1992b,TseytlinET1992a,GasperiniET1992b,BrusteinET1994b,%
BehrndtET1994a,KaloperET1995a},
with particular attention being paid to the effect of the dilaton
field and its self-interaction potential.  However, the roles of the
antisymmetric tensor field and the spatial curvature of the metric
have often been neglected.  Copeland, Lahiri and
Wands~\cite{CopelandET1994b} give the general analytic solutions to
the equations of motion for the antisymmetric tensor field in a
background with non-zero spatial curvature, for the case when the
central charge is zero. In this paper we consider this system with the
a non-zero central charge. Previously, Tseytlin
\cite{Tseytlin1992a} has given several exact solutions. Also Goldwirth
and Perry~\cite{GoldwirthET1993a} use a phase-plane analysis to
describe the cosmological properties of solutions to the equations of
motion of models with a non-zero central charge in a flat FLRW
background.

Many previous authors have exploited the equivalence of the action
written in terms of conformally related metrics. In this paper we show
that there is an additional, formal, equivalence between the equations
of motion for our homogeneous fields in {\em spatially curved\/} FLRW
metrics and those of a scalar field minimally coupled to Einstein
gravity in a {\em spatially flat\/} FLRW metric. We refer to this
system as the shifted FLRW frame, and use it to succinctly describe
the full range of cosmological evolution that is possible within this
model. As the name suggests, it is based on the shifted dilaton
field~\cite{Veneziano1991a,Tseytlin1991a}.  While we cannot give an
analytic solution to the equations of motion for the general case when
the central charge is non-zero, we do find an exact result for a
particular choice of parameters.

\section{Tree-Level String Effective Action}

We take as our starting point the tree-level action
\cite{FradkinET1985a,CallanET1985a,Lovelace1986a}
\begin{equation}
S = {1\over2 \kappa_D^2} \int d^Dx \ \sqrt{-g_D} \ e^{-\phi}
 \left[ R_D + (\nabla\phi)^2 - \Lambda - {1\over12}H^2 \right]  ,
  \label{dimDaction}
\end{equation}
where $g_{ab}$ is the graviton, $\phi$ the dilaton, $H_{abc} =
\partial_{[a}B_{bc]}$ is the antisymmetric tensor field where
lower case roman indices run from $0$ to $D-1$ and the central charge
deficit is denoted by $-\Lambda$. Henceforth we will assume that this
$D$-dimensional theory has undergone compactification, leaving only
four macroscopic dimensions, and that the terms corresponding to
degrees of freedom on the compact metric are held fixed. The real
world may be more complicated, but, in the absence of compelling
reasons for choosing any particular compactification scheme, this
simplification will allow us to examine the dynamics due solely to the
degrees of freedom associated with the macroscopic dimensions. This
allows us to reduce the action to
\begin{equation}
S = {1\over2 \kappa^2} \int d^4x \ \sqrt{-g} \ e^{-\phi}
 \left[ R + (\nabla\phi)^2 - \Lambda - {1\over12}H^2 \right]
  , \label{dim4action1}
\end{equation}
where the spacetime manifold is now four dimensional.

We want to consider a homogeneous and isotropic spacetime, so the
curvature scalar is that of an FLRW universe. The string frame line
element is thus
\begin{equation}
ds^2 = s(\eta)^2 \left[ -n(\eta)^2 d\eta^2 + \frac{1}{1 - kr^2}dr^2
       + r^2 \left( d\theta^2 + \sin^2(\theta)d\phi^2\right)\right]
   \label{line.element}
\end{equation}
where $\eta$ is the conformal time if we set the arbitrary lapse
function $n=1$. Open, flat or closed spatial hypersurfaces correspond
to $k = -1, 0, +1$, respectively. The antisymmetric tensor $H$ has
only one degree of freedom, and can be written
\begin{equation}
H^{\mu\nu\lambda}
 = e^{\phi}\epsilon^{\mu\nu\lambda\kappa}\partial_\kappa \Theta,
   \label{thetadef}
\end{equation}
where $\Theta$ is a pseudo-scalar field. Assuming that the dilaton and
antisymmetric tensor field are homogeneous like our metric,
$H^2=6e^{2\phi}\Theta'^2$, where the prime denotes differentiation
with respect to $\eta$. Furthermore, since the Lagrangian does not
depend on $\Theta$, the corresponding momentum,
\begin{equation}
p_{\Theta} = \frac{\partial {\cal{L}}}{\partial \Theta'}
   = - \frac{ e^{\phi} s^2 \Theta'}{n} = q,
\end{equation}
is conserved and $q$ is a constant. Finally, by adding the total
derivative,
\begin{equation}
-6 \frac{d}{d\eta} \left(\frac{e^{-\phi}s's}{n}\right),
\end{equation}
to the integrand, we obtain
\begin{equation}
S =  \frac{1}{2\kappa^2}\int{ d\eta  \left[ n e^{-\phi}\left(
    -6\frac{s'^{2}}{n^2}   - \frac{\phi'^2}{n^2}s^2
    + 6\frac{\phi's's}{n^2} + 6ks^2   - \Lambda s^4 -
    \frac{q^2}{2s^2}  \right) \right]}.
 \label{Saction3}
\end{equation}

The action is now has a  simpler form, but notice that it includes
the crossed kinetic term $\phi's'$. This term vanishes in the
Einstein frame, which is related to the string frame by the conformal
transformation,
\begin{equation}
s(\eta) = e^{\phi/2}a(\eta),
\end{equation}
where $a$ is the Einstein frame scale factor. In the Einstein frame
the action takes the form
\begin{equation}
S = \frac{1}{2\kappa^2}\int{ dt \ e^{3\alpha}
 \left[ -6 \dot{\alpha}^{2} + \frac{\dot{\phi}^2}{2}
 - U(\alpha,\phi)\right] }
     \label{Eaction3} \ ,
\end{equation}
where for convenience we let $a=e^\alpha$, and choose $n=1/a$ so that
$\eta$ coincides with the Einstein frame  proper time, $t$.

The potential is a function of $\alpha$ and $\phi$,
\begin{equation}
U = \frac{q^2}{2}e^{-2\phi - 6\alpha} + \Lambda e^\phi -
6ke^{-2\alpha} \ .
\end{equation}
However, this potential is simpler when written in
terms of the original string frame scale factor, as it then is a
separable function of $s$ and $\phi$,
\begin{equation}
U = e^{\phi}\left( \Lambda +  \frac{q^2}{s^6} -
\frac{6k}{s^2}\right).
\end{equation}

In fact, the Einstein frame represents only one of infinitely many
different choices of variables which diagonalize the kinetic terms in
the Lagrangian for homogeneous fields. As we shall now see, it is
possible to further simplify the system by choosing alternative
variables which have orthogonal kinetic terms and
respect the symmetry of the potential.

\section{The Shifted  Frame}

While we can simplify the kinetic terms by converting to the Einstein
frame, we do so at the cost of introducing a more complicated
potential. We now introduce a choice of variables that combines the
advantages both of the string frame (separable potential) and the
Einstein frame (no crossed kinetic terms).
Transform $\alpha$ and $\phi$ into a new pair of variables, $r$ and
$\psi$
\begin{equation}
\left( \begin{array}{c}
  \phi \\
  \alpha
\end{array} \right)  =
\left( \begin{array}{cc}
  3  & -3 \\
  -\frac{1}{2} & \frac{3}{2} \end{array} \right)
\left( \begin{array}{c}
  \psi \\
   r
\end{array} \right) \label{shift}
\end{equation}
and choose the  lapse function  to be
\begin{equation}
n = \frac{3}{2}  e^{-\alpha-(\phi/2)}  =
   \frac{3}{2}  e^{-\psi} \ .
\end{equation}

In terms of the transformed variables the action and potential are
\begin{eqnarray}
S &=& \frac{1}{2\kappa^2} \int{dt \ e^{3r}  \left[
   -6 \dot{r}^2 + 2\dot{\psi}^2  - V(\psi) \right]}  \label{Eaction4}, \\
V(\psi) &=& \frac{3}{4}q^2 e^{-6\psi} + \frac{3}{2} \Lambda - 9ke^{-2\psi}.
    \label{Vpsi}
\end{eqnarray}
The potential $V(\psi)$ is plotted for a variety of different
parameter values in Fig.~(1).

Up to a rescaling of the field, this action is identical to that of a
scalar field $\psi$ with potential $V$,  minimally coupled to
Einstein gravity in a spatially flat  universe, with scale factor
$R = e^r$. The equations of motion are
\begin{eqnarray}
&-6\left( \frac{d r}{d T} \right)^2 + 2\left( \frac{d \psi}{d T} \right)^2
    + V(\psi) = 0,&
    \label{shift.constraint} \\
&  \frac{d^2 r}{d T^2}  = -\left( \frac{d \psi}{d T} \right)^2, &
     \label{ddr}\\
& \frac{d^2 \psi}{d T^2} + 3 \frac{d r}{d T}\frac{d \psi}{d T}
   + \frac{1}{4} \frac{d }{d\psi}V(\psi) = 0.
     \label{ddpsi}  &
\end{eqnarray}
The time, $T$, corresponding to our choice of lapse is proportional to
the proper time in the original string frame. It is related to the
proper time in the Einstein frame, $t$, by
\begin{eqnarray}
&& \frac{dT}{dt} = \frac{2}{3}e^{\phi/2} \nonumber \\
&\Rightarrow& t = \frac{3}{2}\int{e^{-\phi/2} dT}. \label{Tt}
\end{eqnarray}

We will refer to this choice of variables as the shifted frame
because, apart from a numerical factor, $r$ coincides with the shifted
dilaton previously
used~\cite{Veneziano1991a,Tseytlin1992a,TseytlinET1992a} to simplify
the equations of motion. Our scalar field is actually the logarithm of
scale factor in the string frame, $\psi\equiv\ln s$, and the shifted
scale factor $R = e^r$ represents the variation of the fields in the
orthogonal direction.  The shifted dilaton, or equivalently $r$,
reflects the symmetries of the underlying string theory better than
the ``renormalized'' field $\phi$~\cite{Veneziano1991a}. For instance,
it remains invariant under the scale-factor duality transformation
$s\to s^{-1}$. However, our action is only invariant under this
transformation if $dV/d\psi=0$, which requires both the spatial
curvature and antisymmetric tensor field to vanish.

This change of variables is not a conformal transformation, which is
an identity between two actions for all field configurations, as the
shifted frame only exists for homogeneous fields.  Also there is no
equivalent choice of variables if we include loop-corrections to the
dilaton coupling. However, if we are going to restrict our attention
to the tree-level action which contains only the dilaton,
antisymmetric tensor and central charge terms with a homogeneous and
isotropic string metric, then working in the shifted frame allows us
to write this system in a particularly simple form.

The practical advantage of working in this shifted frame is that the
spatial curvature of the string metric, $k/s^2$, appears as a term in
the potential $V(\psi)$, and the shifted FLRW metric is always
spatially flat.  The dynamical system in the shifted frame is thus the
basis of almost all inflationary models and its properties are
familiar and well understood. Moreover we will see in the next section
that it is particularly simple to identify the $(+)$ or $(-)$ branches
of pre-big-bang string cosmology~\cite{BrusteinET1994b,KaloperET1995a}
with the contracting or expanding scale factor in the shifted frame.

The qualitative dynamics of our system are strongly influenced by
whether or not the values of $q$, $\Lambda$ and $k$ permit the
inequality $V <0$ to be satisfied. If $\Lambda<0$, then $V < 0$ for
sufficiently large values of $\psi$ but when $\Lambda>0$, a negative
potential region can only exist if $k > 0$.  Then (for $k=1$)
$V(\psi)$ has its minimum value when $\psi = \frac{1}{2}\ln{(|q|/2)}$,
which is negative if $\Lambda |q| < 8$ (see Fig.~(1)).  Thus the
negative potential region exists when one of the following is true:
\begin{equation}
\begin{array}{c}
 \Lambda |q| < 8 \ \mbox{and} \ k=1\\
 \Lambda < 0.
\end{array}
\label{fr.exists}
\end{equation}

In the next section we use the formalism of the shifted frame to
discuss exact solutions to the equations of motion. Analytic solutions
are known to exist when any one of the three terms in $V(\psi)$ is
non-zero, and we show that these can be simply expressed in the shifted
frame.  Secondly, we employ techniques developed for obtaining exact
scalar field cosmologies to derive a new particular solution where all
three terms in the potential are non-zero. In Section 5 we then utilize
these exact solutions as limiting cases of the solutions to the
equations of motion to catalog all the different possible types of
cosmological behavior this model can produce.

\section{Exact Solutions to the Equations of Motion}

Exact solutions to the system of Eqs.~(\ref{shift.constraint})
to~(\ref{ddpsi}) for specific potentials have been analyzed by a
number of authors.
Clearly, the only solutions of relevant to the system considered here
are those where the potential takes the form of Eq.~(\ref{Vpsi}). It
is often useful to parameterize the motion by the field, $\psi$
\cite{Muslimov1990a,SalopekET1990a,Lidsey1991b,Easther1993b,LiddleET1994a}.
Setting $dr/dT = H$ (the ``Hubble parameter'' in the shifted metric),
Eq.~(\ref{ddr}) gives $dH/d\psi = -d\psi/dT$, where the dash now
denotes differentiation with respect to $\psi$. We derive the following
expressions for the potential, scale factor, $r$ and time, $T$:
\begin{eqnarray}
V(\psi) &=& 6H(\psi)^2 - 2 {H(\psi)'}^2,
   \label{Vgenpsi} \\
r(\psi) -r(\psi_0) &=& -\int_{\psi_0}^{\psi}{ \frac{H}{ H'(\psi)}d\psi},
     \label{rgenpsi} \\
T(\psi) -T(\psi_0) &=&
    -\int_{\psi_0}^{\psi}{ \frac{1}{H'(\psi)} d\psi}. \label{Tgenpsi}
\end{eqnarray}
It is convenient to make the extra substitution
\cite{Mitrinovitch1937a,Mitrinovitch1937b,Muslimov1990a}:
\begin{eqnarray}
x &=& \sqrt{3}\psi \ , \\
f(x) &=& \sqrt{\frac{|V|}{6}} \ ,\\
y^2(x) &=& 1 - \frac{V}{6H^2}  \ .  \label{Hgenx}
\end{eqnarray}
There is a sign ambiguity of $H$ corresponding to
a time reversal, which
transforms an expanding solution in the shifted frame into a
contracting one. In terms of $y$, Eq.~(\ref{Vgenpsi}) becomes
\begin{equation}
y y_x = (1 - y^2)\left( y - \frac{f_x}{f}\right) \ ,
   \label{ygenx}
\end{equation}
where $y_x$ denotes $y$ differentiated with respect to $x$. When
$f_x/f$ is constant (or zero) this equation can be integrated
immediately.  Fortunately these special cases correspond to a potential
that is either a constant or a single exponential term, which is
precisely what we need to discuss the asymptotic solutions in Section 5.

\subsection{Pure Dilaton Cosmology}

Firstly, note that in the absence of a central charge, antisymmetric
tensor field and spatial curvature in the string frame then $V=0$ and
we have the standard result for a massless scalar field in the shifted
frame:
\begin{eqnarray}
R & = & R_0 \left| {T\over T_0} \right|^{1/3} \ ,\\
\psi - \psi_0 & = & \pm {1\over\sqrt{3}} \ln \left| {T\over T_0} \right| \ .
\end{eqnarray}
In terms of the string frame scale factor and dilaton this is the usual
pure dilaton cosmology,
\begin{eqnarray}
s & = & s_0 \left| {T\over T_0} \right|^{\pm1/\sqrt{3}} \ , \\
\phi - \phi_0 & = & \left( \pm\sqrt{3}-1 \right)
 \ln \left| {T\over T_0} \right| \ .
\end{eqnarray}
The choice of signs in the above equations correspond to an
increasing or decreasing dilaton. In addition there are two branches
corresponding to $T$ less than or greater than zero, denoted as the
$(+)$ and $(-)$ branches by Brustein and
Veneziano~\cite{BrusteinET1994b}.  We see that in the shifted frame
these two branches correspond simply to a contracting or expanding
scale factor, $R$, respectively.\footnote{This can seen from equation~(4)
of~\cite{BrusteinET1994b} where the choice of $\pm$ sign coincides
with the sign of $-\dot{r}$.}
Here the $(+)$ branch approaches a singularity at $T=0$, while the
$(-)$ branch starts from the singularity at $T=0$.

\subsection{Dilaton Cosmology with a Central Charge}

When $\psi \gg 0$, $V(\psi)$ is dominated by the central charge term
$\Lambda$ and we  approximate the potential by $V = 3\Lambda/2$.
For this case the solution is straightforward, since $f_x = 0$ and
\begin{equation}
y(x) = \left\{
\begin{array}{lc}
\tanh{(x-x_0)} & {\rm for\ }\Lambda>0 ,\\
\coth{(x-x_0)} & {\rm for\ }\Lambda<0 .
\end{array}
 \right.
 \label{Ay1}
\end{equation}

When $\Lambda$ is positive, $f= \sqrt{\Lambda}/{2}$ and
\begin{eqnarray}
R(\psi) &=& R_0
  \left\{ \sinh{ \left[\sqrt{3} (\psi - \psi_0 ) \right]}
\right\}^{-1/3},    \label{AR1pos}\\
T(\psi) - T_0 &=& \mp \frac{2}{3\sqrt{\Lambda}} \ln{ \left\{
    \tanh{ \left[ \frac{\sqrt{3}}{2} (\psi - \psi_0 ) \right]} \right\} }.
 \label{AT1pos}
\end{eqnarray}
This is the general solution for a massless field plus cosmological
constant in a flat FLRW metric~\cite{SchunckET1994a}. The choice of
upper or lower sign reflects whether we choose the $(+)$ or $(-)$
branch corresponding to the contracting or expanding solutions,
respectively, in the shifted frame. The string frame scale factor and
dilaton are~\cite{KaloperET1995a}
\begin{eqnarray}
s & = & s_0 \left[ \tanh \left( {3\sqrt{\Lambda}\over2} |T-T_0|
\right) \right]^{\pm1/\sqrt{3}} \ , \\
e^{\phi-\phi_0} & = & \frac{
 \left[
  \tanh \left( {3\sqrt{\Lambda}\over4} |T-T_0| \right)
   \right]^{\pm\sqrt{3}}}{
  \sinh \left( {3\sqrt{\Lambda}\over2} |T-T_0| \right) } \ .
\end{eqnarray}

The corresponding solution with $\Lambda <0$ has
$f=\sqrt{|\Lambda|}/{2}$ and
\begin{eqnarray}
R(\psi) &=& R_0
  \left\{ \cosh{ \left[ \sqrt{3} (\psi - \psi_0 ) \right]}
\right\}^{-1/3},    \label{AR1neg}\\
|T(\psi) - T_0| &=&  \frac{4}{3\sqrt{\Lambda}} \tan^{-1}{ \left\{
    \exp{ \left[ \sqrt{3} (\psi - \psi_0 ) \right]} \right\} }.
 \label{AT1neg}
\end{eqnarray}
There is no choice of $(+)$ or $(-)$ branches, as every solution for
$R(\psi)$ starts expanding [$(-)$ branch], turns around when
$\psi=\psi_0$ and recollapses [$(+)$ branch].  Written in terms of the
string frame variables we have
\begin{eqnarray}
s & = & s_0
 \left[ \tan \left( {3\sqrt{|\Lambda|}\over4} (T-T_0)
       \right) \right]^{\pm1/\sqrt{3}}\ , \\
e^{\phi-\phi_0} & = &
 \frac{
       \left[ \tan \left( {3\sqrt{|\Lambda|}\over4} (T-T_0)
                           \right) \right]^{\pm\sqrt{3}}
 }{ \cos \left( {3\sqrt{|\Lambda|}\over2} (T-T_0) \right) }
\ .
\end{eqnarray}
Note that the $\Lambda < 0$ solution has a finite lifetime, as
$|T-T_0|$ is bounded between $0$ and $2\pi/3\sqrt{|\Lambda|}$.  The
string scale factor is monotonic, and increases if we take $T > T_0$
and decreases otherwise.

\subsection{Dilaton and Antisymmetric Tensor Field}

When $\psi \ll 0$, the potential has the form $V(\psi) =
3q^2e^{-6\psi}/4$, so $f_x/f = -\sqrt{3}$.  For this case we give
$y$ parametrically  as
\begin{eqnarray}
\psi(y) - \psi_0 &=& -\frac{1}{4}
  \ln{ \left[ \left( \frac{1-y}{1+y} \right)^{1/ \sqrt{3} }
  \frac{(\sqrt{3} + y)^2}{1-y^2} \right] }  \label{Ay2}  \\
R(y) &=& R_0  \left[ \left( \frac{1-y}{1+y} \right)^{\sqrt{3}}
     \frac{(\sqrt{3} + y)^2}{1-y^2}   \right]^{1/12}.  \label{AR2}
\end{eqnarray}
A scalar field with an exponential potential is the basis of power-law
inflation~\cite{LucchinET1985a,LucchinET1985b} and is known to have an
exact solution~\cite{SalopekET1990a}.  The situation in the shifted
frame is not analogous to power-law inflation, due to the steepness of
the potential. In particular, we find that the value of $\psi$ is
always bounded below, whereas the exponential potentials which drive
power-law inflation admit solutions where the field is a monotonic
function of the time. For $|f_x/f|>1$ the value of $\psi$ cannot
decrease indefinitely, irrespective of the initial conditions.
The minimum value occurs when $y=0$, while at early or late times, as
$y\to\pm1$, we have $\psi\to+\infty$. Thus the string frame scale
factor, $s=e^\psi$, always has a minimum value. This behavior is also
seen in the exact solution given, in rather different form, by
Copeland, Lahiri and Wands~\cite{CopelandET1994b} for this system with
$\Lambda=0$.

\subsection{Special Case: Particular Solution with $k=+1$}

We have found a new exact solution to the equations of motion, for a
case where all the terms in the potential Eq.~(\ref{Vpsi}) are
non-zero. This is generated by choosing
\begin{equation}
H(\psi) = \pm \left( A - B e^{-2\psi} \right)^{3/2},  \label{Hexact1}
\end{equation}
where $A$ and $B$ are both positive. The plus and minus signs
correspond to increasing and decreasing $r$, respectively, with $H=0$
when $\psi=\psi_0\equiv(1/2)\ln(B/A)$.  It is straightforward to write
down the potential
\begin{equation}
V(\psi) = 6A^3 - 18A^2 B e^{-2\psi} + 12B^3 e^{-6\psi}. \label{Vpsi1}
\end{equation}
If either $A$ or $B$ vanishes then the potential reduces to one of the
special cases we have already considered, so we assume that they are
both non-zero. By comparison with Eq.~(\ref{Vpsi}), we see that this provides
us with a non-trivial solution to the equations of motion when
$\Lambda > 0$ and
\begin{eqnarray}
A = \left(\frac{\Lambda}{4}\right)^{1/3} &,&
B = \frac{1}{2A^2} \\
q^2 \Lambda^2 &=& 32.
\end{eqnarray}

Performing the integrals in Eqs~(\ref{rgenpsi}) and~(\ref{Tgenpsi}) yields
\begin{eqnarray}
r(\psi) -r_0 &=&  \frac{1}{3} ( \psi - \psi_0 )
     - \frac{1}{6} \left( e^{2 (\psi - \psi_0)} -1 \right) \label{rexact1}\\
T(\psi)-  T_0 &=& \mp \frac{1}{3\sqrt{\Lambda}}\left[
   e^{\psi - \psi_0}   \sqrt{e^{2(\psi - \psi_0)} -1}  +
 \ln{\left( e^{\psi - \psi_0}  +  \sqrt{e^{2(\psi - \psi_0)} -1} \right)}
 \right]
  \label{Texact1}
\end{eqnarray}
This solution is displayed in Fig.~(2). Notice that
$\psi_0$ is the minimum value attained by $\psi$, and quantities with
the subscript $0$ refer to their value at $\psi = \psi_0$. The
maximum value of $r$ is  $r_0$.  The upper sign in the expression for $T$
corresponds to the expanding phase ($T < T_0$), and the lower sign to
the contracting phase ($T > T_0$).  This solution thus interpolates
between the $(-)$ branch and the $(+)$ branch.

The ambition of many studies in string cosmology has been to show
whether a non-singular universe can be found as a solution to the
equations of
motion~\cite{BrusteinET1994b,BehrndtET1994a,%
KaloperET1995a,AngelantonjET1994a}.
Because the lifetime in the shifted frame, and thus in the string
frame, for our solution is infinite, it might appear to be just such a
non-singular cosmology.  However, the dilaton,
\begin{equation}
\phi - \phi_0 =  2(\psi - \psi_0)  + \frac{1}{2} \left( e^{2(\psi -
\psi_0)} - 1 \right) ,
 \label{phiexact1}
\end{equation}
becomes arbitrarily large when $|T| \to \infty$. This is the strong
coupling limit of the string theory, so the tree-level action from
which our solution is derived becomes unreliable. This is a
consequence of the change at $T_0$ from the $(-)$ to $(+)$ branches,
rather than changing from $(+)$ to $(-)$, as envisaged in the pre-Big
Bang scenario~\cite{BrusteinET1994b}.

The Einstein frame scale factor, $a = e^\alpha$ is given by
\begin{equation}
\alpha - \alpha_0 =  - \frac{1}{4} \left( e^{2(\psi - \psi_0)} -1
\right) ,  \label{alphaexact1}
\end{equation}
The time in this frame is given by Eq.~(\ref{Tt}), so
\begin{equation}
t(\psi) - t_0 = \pm \frac{e^{\alpha_0 + 1/4}}{2\sqrt{2}} \int_z^1
     {\frac{e^{-1/z'}}{z'^{3/2}\sqrt{1-z'}} dz'} \label{texact1}
\end{equation}
where we have made the additional substitution $z = e^{-2(\psi
-\psi_0)}$.  Evaluating this integral with the lower limit $z=0$ shows
that the time between the ``Big Bang'', when $a=0$, and the time $t_0$
when the universe attains its maximum size is finite in the Einstein
frame.

Finally, we remark that this particular solution is unstable to small
perturbations. In the shifted frame, our solution corresponds to the
critical case where the $\psi$ field reaches an infinite value in an
infinite time with vanishing velocity. If it rolled more slowly it would
eventually be reflected back towards the minimum of the potential,
whereas a faster evolution would see it become infinite in a finite
time.

\subsection{Special Case: Static Solution with $k=+1$}

Finally, there is a particular solution when the $\psi$ sits in the
minimum of its potential, if this minimum value is non-negative.
We can therefore find the result, with $\dot\psi=0$, first given by
Tseytlin~\cite{Tseytlin1992a},
\begin{eqnarray}
\psi &=& \frac{1}{2}\ln{\frac{|q|}{2}},  \label{psiexact2} \\
r(T) - r(0) &=& \pm \sqrt{\frac{2}{|q|} - \frac{\Lambda}{4}} \ T ,
   \label{rexact2}
\end{eqnarray}
when $k=+1$ and $\Lambda|q|\geq8$.
Since $s = e^\psi$, the string frame scale factor, $s$, is a constant.
The dilaton,
\begin{equation}
\phi - \phi_0 = \mp 3\sqrt{\frac{2}{|q|} - \frac{\Lambda}{4}} \ T ,
\end{equation}
is linear with respect to the string frame time.  The Einstein frame
scale factor, $a$, is monotonic and proportional to the Einstein frame
time, $t$. Like our previous particular solution, the lifetime in the
string frame is infinite, but the dilaton becomes large at either
early or late times, here depending on whether we are on the $(+)$ or
$(-)$ branch, rendering the tree-level action invalid. There is a
corresponding singularity in the Einstein frame when the scale factor
becomes zero.

This $(-)$ branch solution is stable at late times, as perturbations of
$\psi$ about the minimum are damped by the expansion of the shifted
frame. Conversely, the $(+)$ branch solution is unstable at late
times, but is the general solution at early times.

\section{Cosmological Behavior for the General Case}

The exact solutions examined in the previous section only apply to a
small portion of the full parameter space. However, using the shifted
frame we can give a qualitative account of the properties of the
general solution to the equations of motion.

For large negative and decreasing $\psi$ with $q \neq 0$, the system
must eventually evolve into a region where Eqs~(\ref{Ay2})
to~(\ref{AR2}) accurately describe the motion. This shows that $\psi$
cannot decrease to arbitrarily large negative values. This immediately
establishes that the string frame scale factor, $s=e^\psi$ always has
a non-zero lower bound in the presence of an antisymmetric tensor
field.  Conversely, we will show that $\psi$ always reaches
arbitrarily large values at early and/or late times, except in the
particular static solution of Eq.~(\ref{psiexact2}). The evolution of
the string frame scale factor can be quite complicated but in the
shifted frame the evolution is straightforward.

If the scale factor in the shifted frame is growing, the energy
density must decrease and the field will eventually evolve towards the
minimum of its potential.  This naturally splits the analysis into two
sub-cases, depending on whether or not the values of $q$, $\Lambda$
and $k$ admit a negative potential region and we treat them
separately.

\subsection{The Motion with a Negative Potential Region}

We showed in Section~3 that a negative potential region
exists whenever $\Lambda < 0$ or $k > 0$ and $\Lambda |q| < 8$. By
examining the constraint, Eq.~(\ref{shift.constraint}), we can catalog
the possible extrema of $r(T)$ and $\psi(T)$.

\begin{itemize}

\item{ $\dot{r} = 0$, $\dot{\psi} \neq 0$ }

We see that for $\dot{r} =0$ we must have $V(\psi) < 0$, so branch
changing between contracting and expanding solutions can only occur in
the negative potential region.  In addition, Eq.~(\ref{ddr}) implies that all
turning points of $r(T)$ are maxima, so all branch changes must be
{}from the $(-)$ to the $(+)$ branch. Therefore $r$ has at most only one
turning point.

\item{ $\dot{\psi} = 0$, $\dot{r} \neq 0$  }

Extrema of $\psi(T)$ can only occur outside the negative potential
region and reflect $\psi$ back towards the minimum of the potential.

\item{\bf  $\dot{r} = \dot{\psi} = 0$}

This special case can only occur on the boundary of the negative
potential region. Again, the value of $r(T)$ is a maximum, and the
trajectory is reflected back towards $V<0$. Our exact solution,
Eqs~(\ref{rexact1}) to~(\ref{Texact1}), exhibits this type of extremum.

\end{itemize}

When $\dot{r} >0$, even if the field $\psi$ is evolving away from the
negative potential region, the frictional damping will force $\psi$
towards the minimum of the potential. Eventually the energy density in
the shifted frame (kinetic plus potential energy of $\psi$) reaches
zero, leading to a turning point for $r(T)$.
In the contracting phase the energy density increases. The presence of
the antisymmetric tensor field will ensure that $\psi$ is
always reflected back from large negative values towards the minimum
of the potential.
However, at large $\psi$ the potential energy tends towards a finite
value, $3\Lambda/2$. Once the total energy density exceeds this value,
it must continue to increase if $\dot{r}<0$ and $\psi$ will escape to
infinity. Similarly, extrapolating back in time, we find that $\psi$
must always originate at infinity.

For $\Lambda>0$ (and thus $k=+1$) the field $\psi$ may oscillate about
the minimum of the potential many times both during the expanding and
contracting phases, as shown in Fig.~(3). If
$\Lambda\leq0$ the energy density always exceeds $3\Lambda/2$ and thus
the field escapes to infinity without passing through a local maximum.

Our particular solution, Eqs~(\ref{rexact1}) and~(\ref{Texact1}),
corresponds to the critical case where the asymptotic energy density
is exactly $3\Lambda/2$ and $\psi$ reaches infinity with zero kinetic
energy as $|T|\to\infty$.  Such a late (or early) time solution exists
for any choice of parameters (when $\Lambda>0$), but our exact
solution with $q^2\Lambda^2=32$ is the special case where the turning
points for $r$ and $\psi$ coincide and the evolution is symmetrical
about $T_0$.

For $k=+1$ the Einstein frame scale factor may possess both local
maxima and local minima, which implies that the string matter, after
transformation to the Einstein frame, does not satisfy the strong
energy condition. This behavior can be seen both by numerically
integrating the full equations of motion, or by considering the
equation of motion for $\ddot{\alpha}$.  However, if a negative
potential region exists there is always an upper bound on the Einstein
frame scale factor. Conversely, as we shall see in the next section,
if the potential is everywhere non-negative then the Einstein frame
scale factor is monotonic.

\subsection{The Motion without a Negative Potential Region}

In this case $V \geq 0$ at all points. {}From the constraint equation,
Eq.~(\ref{shift.constraint}), $\dot{r} = 0$ requires both $V$ and
$\dot\psi$ to be zero, which is a special case of the static solution
given in Eqs.~(\ref{psiexact2}) and~(\ref{rexact2}). Otherwise,
without a negative potential region $r$ is monotonic, and therefore
there are no branch changing solutions.

Turning points in $\psi$ will still occur. However, for $k \leq 0$ the
potential is a decreasing, monotonic function of $\psi$ and any
extremum will be a global minimum as there can be no further turning
points. In this case $\psi$ becomes infinitely large at both early and
late times, as illustrated by Fig.~(4).

For $k=+1$, the potential has a minimum value at $\psi =
1/2\ln{|q|/2}$. When $r$ is increasing [$(-)$ branch], the field
oscillates with decreasing amplitude about this minimum, and the
static solution given in Eqs.~(\ref{psiexact2}) and~(\ref{rexact2}) is
a stable attractor at late times. Note that closed models therefore
can escape recollapse in the Einstein frame if $\Lambda|q|\geq8$. This
type of motion is shown in Fig.~(5). The $(+)$ branch is
simply the time reversed solution, and so is unstable at late times.

When $k=-1$ the behavior of this system at late times can be
probed using the slow rolling approximation. The criteria for the
validity of this approximation are that the potential is much larger
than its first and second derivatives, which holds well when $V
\approx 3\Lambda/2$ and $V'(\psi) \approx 18ke^{-2\psi}$. Dropping the
appropriate terms from Eqs~(\ref{shift.constraint}) and~(\ref{ddpsi}),
we integrate the approximate system to obtain the following asymptotic
solution for large $\psi$,
\begin{eqnarray}
r &\to& \frac{\sqrt{\Lambda}}{2} \ T \\
\psi &\to& \frac{1}{2} \ln{\left( \frac{6}{\sqrt{\Lambda}}\ T \right) }
\end{eqnarray}
At late times therefore, $r \gg \psi$, and thus, from
Eq.~(\ref{shift}), $\phi\to-\infty$ and $a\to\infty$. A similar
analysis will show that the late time behavior for $\Lambda,|q| >0$
and $k=0$ is similar to that for $k=-1$.

\section{Conclusions}

We have succeeded in describing the full range of cosmological
evolution that can be found for the string motivated action,
Eq.(\ref{dim4action1}), containing the dilaton, central charge
and antisymmetric tensor field, with a homogeneous and isotropic but
spatially curved metric. We have done this by showing that this system
is formally equivalent to a self-interacting scalar field, $\psi$,
minimally coupled to Einstein gravity in a spatially flat FLRW metric,
and by using this shifted frame to understand the cosmological
evolution.  The parameters of the theory determine the form of the
self-interaction potential, $V(\psi)$.

If the potential for the scalar field in the shifted frame is positive
definite then the generic evolution of the shifted frame scale factor
is monotonic and with a semi-infinite lifetime. Solutions either start
or end at a singularity where the scale factor vanishes. Monotonically
contracting or expanding solutions correspond to the $(+)$ or $(-)$
branches, respectively, of the pre-big-bang
scenario~\cite{GasperiniET1992b,BrusteinET1994b,KaloperET1995a}. If
$V<0$, then a turning point is possible, but this is always a maximum
corresponding to a change from the $(-)$ to $(+)$ branch. These
general conclusions will remain valid for any potential
$V(\psi)$. Generic solutions to this system are singular. Only
exceptional cases, for which we have analytic solutions, have an infinite
lifetime in the string frame. However, the dilaton always diverges at
early and/or late times, taking the solution into the strong
coupling regime.

In the string or Einstein frames the solutions exhibit a
diverse range of behavior, depending on both the curvature of the
spatial hypersurfaces of the background spacetime (described by $k$)
and the other parameter values, $\Lambda$ and $q$. As long as the
antisymmetric tensor field is non-zero, the string frame scale factor
is always bounded from below \cite{CopelandET1994b}.

For all values of $k$, including the case with positive curvature when
$\Lambda > 8/|q|$, there are choices of the parameter values for which
the Einstein frame scale factor expands indefinitely from an initial
singularity. If $k=+1$ and $\Lambda < 8/|q|$ the Einstein frame scale
factor can pass through several local maxima and minima, but the
lifetime of the universe is finite. If $\Lambda < 0$, then the
Einstein frame scale factor always has a finite maximum value,
irrespective of the values of $k$ and $q$.

Each term in the action we have considered turns out to play an
important role at different stages in the cosmological evolution.
Consequently, we have found new types of behavior not seen in previous
studies which omit one or more of the terms. By the same token, our
own conclusions may be sensitive to the inclusion of further terms in
the action. Nonetheless, the absence of solutions which interpolate
between weak coupling regimes rules out the possibility of
successfully implementing the pre-Big Bang
scenario~\cite{GasperiniET1992b} in our system. This complements the
work of Kaloper, Madden and Olive~\cite{KaloperET1995a} who reach
similar conclusions considering the effect of an explicit potential
for the dilaton and loop corrections to the dilaton coupling, but
without an antisymmetric tensor field or spatial curvature. All
possible solutions to our equations of motion contain phases where the
coupling becomes strong. Therefore this tree-level limit of the full
string theory predicts its own downfall, where higher-order
corrections cannot be neglected.

\section*{Acknowledgements}

KM acknowledges the British Council for a travel grant to visit Sussex
University, where this work was begun, and RE would like to thank
Sussex University for its hospitality while this work was completed.
RE is supported by a JSPS post-doctoral fellowship, and the
Grant-in-Aid for JSPS fellows (0694194). DW is a PPARC post-doctoral
research fellow.  The authors would like to thank Ed Copeland and Juan
Garc\'\i a-Bellido for useful discussions. This work was partially
supported by the Grant-in-Aid for Scientific Research Fund of the
Ministry of Education, Science and Culture, Japan (06302021 and 06640412).

\newpage

\section*{Figure Captions}

\mbox{}

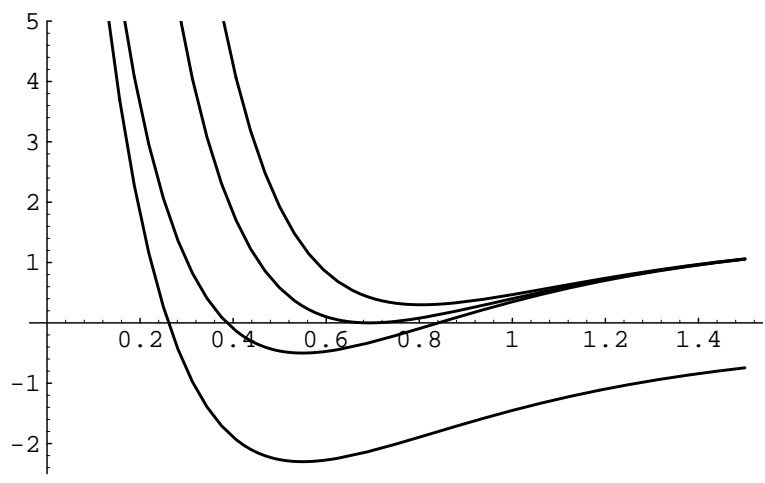
\noindent Figure 1: The potential $V(\psi)$ is shown for a
variety of parameter values. In ascending order, the plots correspond
to $k=1$, $\Lambda = -0.2$ and $q=6$, and $\Lambda =1$, $k=1$ with $q
= 6$, $8$ and $10$.

\medskip
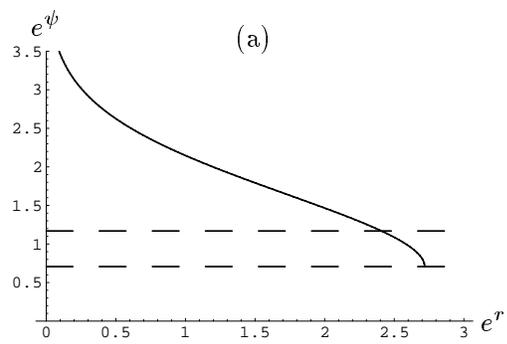
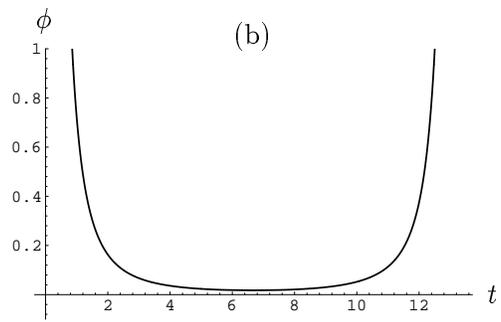
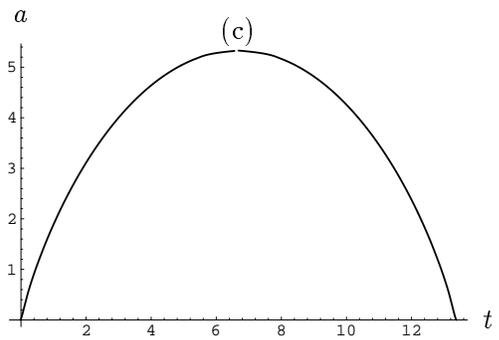
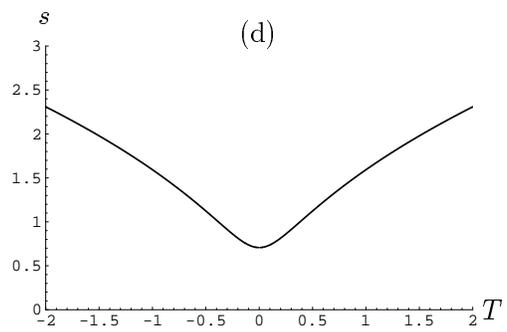
\noindent Figure 2:
This figure displays the exact solution to the string equations of
motion given by Eqs~(\ref{alphaexact1}) and~(\ref{phiexact1}) for the
parameter values $A=1$ ($\Lambda = 4$ and $q^2 = 2$) and $r_0 = 1$. In
(a) the motion in the shifted frame is shown. The path is reflected at
the boundary of the negative potential region, which lies between the
dashed horizontal lines. The evolution of the dilaton $\phi$ is
plotted in (b) and the Einstein frame scale factor is plotted in (c).
The coordinate time, $t$, in the Einstein frame runs over a finite
interval, but in the string frame $T$ runs from $-\infty$ to
$+\infty$. In (d), we plot the string frame scale factor, $s$, when it
is near its minimum value.

\medskip
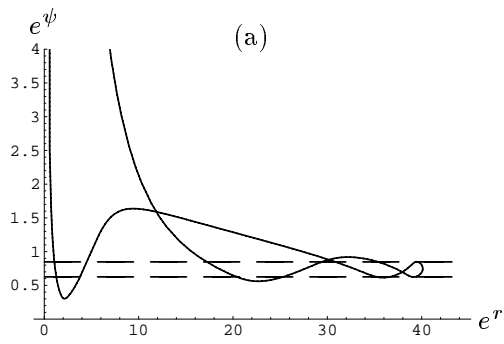
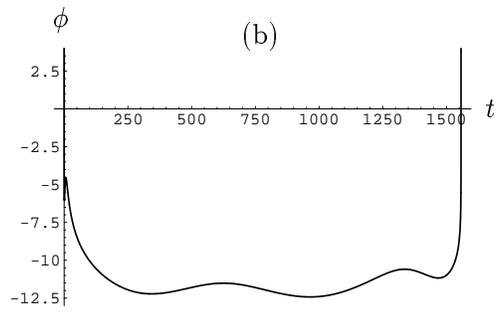
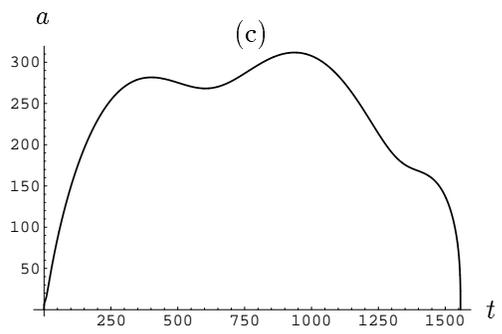
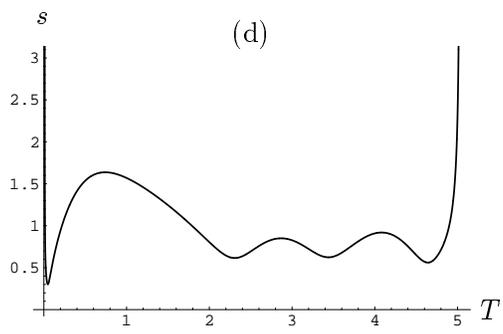
\noindent Figure 3:
Numerical solution of the equations of motion
with $q=1$, $\Lambda = 7$ and $k=1$ is plotted here for the initial data
$r =1$, $\psi = -1$, $\dot{r} = 10$ with $\dot{\psi}$ chosen to
satisfy the constraint. In (a) the motion in the shifted  frame is
shown. The boundary of the negative potential region has also been
plotted (the two horizontal lines) and the oscillations around it
can be clearly seen. The evolution of the dilaton is shown in (b),
while plots (c) and (d) depict the Einstein and string frame scale
factors. Note that while this model is singular (when $a=0$), the
Einstein frame scale factor can have several phases of expansion and
contraction.

 \medskip
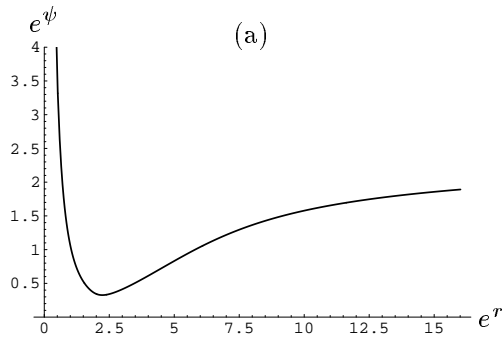
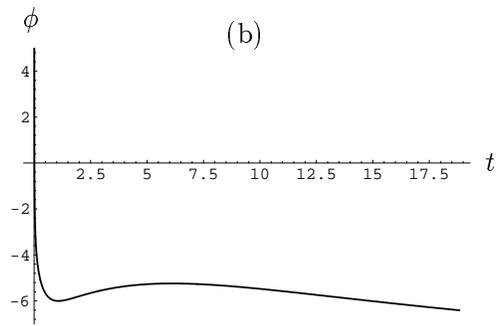
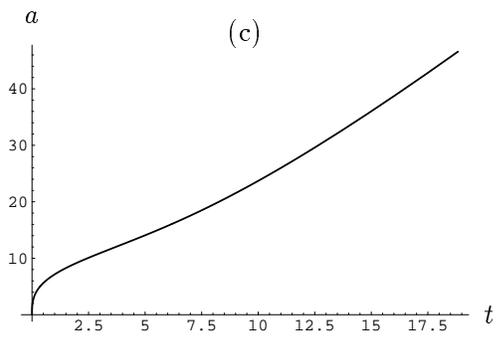
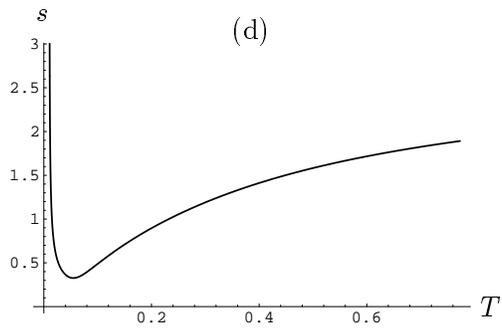
\noindent Figure 4:
The plot shows the solution  with
$\Lambda=9$ and $k=-1$, and the other parameters are the same as for
the plot in Fig.~(3). Since $k= -1$, the field $\psi$
does not oscillate. At late times $r \rightarrow \infty$ and $\psi
\rightarrow -\infty$. The dilaton (b) decreases indefinitely, while
the Einstein frame scale factor (c) expands without limit. The string
frame scale factor (d) is initially infinite, and is proportional to
$T$ at late times.

 \medskip
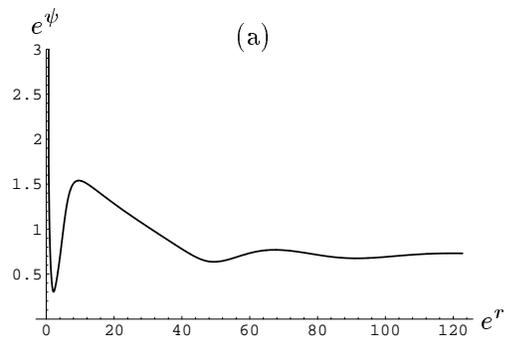
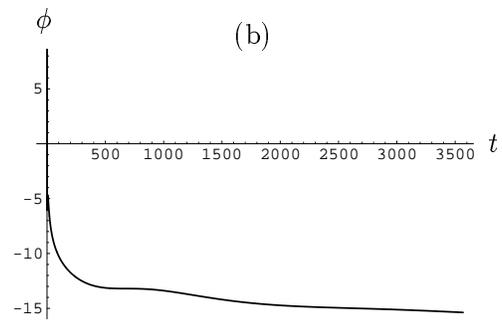
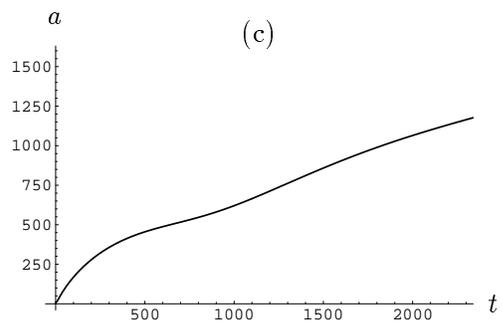
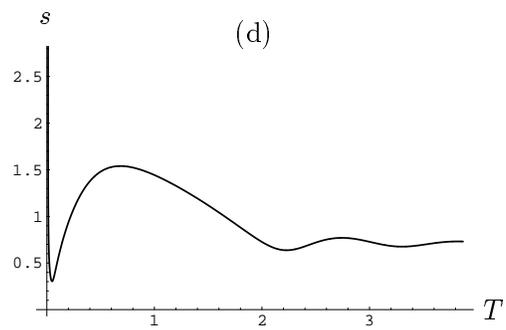
\noindent Figure 5:
This figure shows the solution to the
equations of motion when the parameter values are the same as those in
Fig.~(3), except for setting $\Lambda=9$. In this case
there is no negative potential region as $\Lambda |q| >8$. Since
$k=1$, the field $\psi$ oscillates about the value $e^{-2\psi} =
2/|q|$ and at late times the solution tends towards that given by
Eqs~(\ref{psiexact2}) and~(\ref{rexact2}). This can be observed in the
plots of the dilaton (b) and the Einstein frame scale factor (c), as
well as the string frame scale factor (d) which is asymptotically
constant.

\end{document}